\documentclass[aps,prl,twocolumn,superscriptaddress,floats,epsfig,showpacs]{revtex4}

\usepackage{pstricks,pst-grad,color,graphics}
\usepackage{latexsym}
\usepackage{amsmath}
\usepackage{amssymb}
\usepackage{enumerate}
\usepackage{bbm}
\usepackage{psfrag}
\usepackage{epsfig}

\newcommand{\be}{\begin{equation}}
\newcommand{\ee}{\end{equation}}
\newcommand{\BE}{\begin{equation}}
\newcommand{\EE}{\end{equation}}
\newcommand{\eea}{\end{eqnarray}}
\newcommand{\bea}{\begin{eqnarray}}

\newcommand{\mean}[1]{\ensuremath{\langle{#1}\rangle}}

\newcommand{\eins}{\ensuremath{\mathbbm 1}}
\newcommand{\Eins}{\ensuremath{\mathbbm 1}}

\newcommand{\WW}{\ensuremath{\mathcal{W}}}

\newcommand{\NN}{\ensuremath{\mathcal{N}}}

\newcommand{\ketbra}[1]{\ensuremath{| #1 \rangle \langle #1 |}}

\newcommand{\ket}[1]{\ensuremath{|#1\rangle}}

\newcommand{\bra}[1]{\ensuremath{\langle#1|}}

\newcommand{\braket}[2]{\ensuremath{\langle #1|#2\rangle}}

\newcommand{\kommentar}[1]{}

\def\tr{{\rm tr}}
\def\Emin{\varepsilon}
\def\Rl{{\mathbb R}}
\def\Eh{\widehat E}
\def\idty{\Eins}

\begin{document}
\title{Estimating entanglement measures in experiments}

\author{O. G\"uhne}
\affiliation{Institut f\"ur Quantenoptik und Quanteninformation,
\"Osterreichische Akademie der Wissenschaften,
A-6020 Innsbruck, Austria}
\author{M. Reimpell}
\affiliation{Institut f\"ur Mathematische Physik, 
Technische Universit\"at Braunschweig, 
Mendelssohnstra{\ss}e 3, D-38106 Braunschweig, Germany}
\author{R.F. Werner}
\affiliation{Institut f\"ur Mathematische Physik, 
Technische Universit\"at Braunschweig, 
Mendelssohnstra{\ss}e 3, D-38106 Braunschweig, Germany}

\begin{abstract}
We present a method to estimate entanglement measures in experiments. 
We show how a lower bound on a generic entanglement 
measure can be derived from the measured expectation values of any 
finite collection of entanglement witnesses. 
Hence witness measurements are given a quantitative meaning without 
the need of further experimental data.
We apply our results to a recent multi-photon 
experiment [M.~Bourennane {\it et al.}, Phys. Rev. Lett. {\bf 92}, 
087902 (2004)], giving bounds on the entanglement of formation and 
the geometric measure of entanglement in this experiment.
\end{abstract}
\pacs{03.65.-w, 03.65.Ud, 03.67.-a}
\maketitle


{\it Introduction ---}
Deciding whether or not a given state is entangled is one of the
basic tasks of quantum information theory. In principle, one can
determine the full quantum state via state tomography, and apply 
some separability criteria afterwards. However, the tomography 
requires an effort which is growing exponentially with the number 
of parties. For practical implementations, it is therefore
highly desirable to verify entanglement on the basis of only a few,
maybe only one measurement. 
Entanglement witnesses \cite{horo} are just the observables for this 
purpose: by definition they have positive expectation on every 
separable state, so when a negative expectation is found in some 
state, it must be entangled. Consequently, entanglement witnesses 
have been used in many experiments \cite{witexp, mohamed, hartmut}, 
and their theory is far developed \cite{wittheo, brandao, nlew}.

Besides the mere detection, the quantification of entanglement 
is an even more challenging problem in the field. Here one aims
at characterizing the amount of entanglement by so-called entanglement 
measures. Many entanglement measures have been 
introduced for this purpose \cite{plenio}. But even if a quantum state is fully
known, the computation of  given entanglement measure is often not 
straightforward. Needless to say that the efficient determination of 
an entanglement measure in experiments is even more complicated.

In this paper we present a method to estimate entanglement measures
in experiments. We show that entanglement witnesses cannot only be 
used for the detection of entanglement,  but also for its quantification:
any measured negative expectation value of a witness can be 
turned into a nontrivial lower bound on a generic entanglement measure. 
Hence, if witnesses are already used for entanglement detection, 
the estimation of an entanglement measure requires no extra 
experimental effort. 
We describe the procedures for computing such bounds in detail for 
entanglement of formation \cite{eof} and the geometric measure of 
entanglement \cite{wei}. 
Our method can not only be applied to the measurement of a single 
witness, but extends to incomplete tomography in general: for any 
finite set of measured expectation values we characterize the best 
possible lower bound on any convex entanglement measure 
(or, more generally, any convex figure of merit) consistent with 
these expectations. 
Finally, we apply our results to a recent multi-photon 
experiment~\cite{mohamed}. 

The theoretical context of our method is the theory of Legendre 
transforms (also called Fenchel transforms or conjugate functions) 
\cite{rockafellar}. This method has already been used to characterize 
additivity properties of entanglement measures~\cite{lit:koenraad}. 
The question how to estimate the entanglement when only partial 
knowledge is given was, to our knowledge, first addressed 
in Ref.~\cite{horoneu}. Bounds on some entanglement measures 
from special Bell inequalities or entanglement witnesses have been
obtained in Refs.~\cite{wolf, brandao}, and methods to estimate 
measures in experiments by making measurements on several 
copies of a state have been discussed in Ref.~\cite{mintert}. 
While finishing this paper, we learned that similar ideas and conclusions, 
illustrated with a discussion of a complementary choice of entanglement 
measures, are reached in a paper 
by Eisert, Brand\~ao and Audenaert \cite{eisert}.

{\it Main idea of the estimation ---} Let us consider $n$ witness operators 
(or indeed any hermitian operators \cite{remark1}), 
$\WW_1,\ldots,\WW_n$ on the same Hilbert space, 
and some entanglement measure $E$, assigning to every density operator
$\rho$ a numerical value $E(\rho)$ characterizing its entanglement.
We assume for the moment only that $\rho\mapsto E(\rho)$ is convex
and continuous. Suppose now that, for some state $\rho$, we have
measured the expectations of the $\WW_k$, i.e., that we are given the
real numbers $w_k=\tr(\rho \WW_k)$ for $k=1,\ldots,n$. On the basis of
these numbers we would like to calculate a lower bound on $E(\rho)$
or, more precisely, the best lower bound
\begin{equation}
\label{Emin}
\Emin(w_1,\ldots,w_n)
    =\inf_\rho \left\{E(\rho)\,|\,\tr(\rho \WW_k)=w_k\right\},
\end{equation}
where the infimum is understood as the infimum over all states
compatible with the 
data $w_k=\tr(\rho \WW_k).$

\begin{figure}[t]
\setlength{\unitlength}{0.1\columnwidth}
\begin{picture}(10,6)
\thinlines
\put(1,0){\includegraphics[width=0.8\columnwidth]{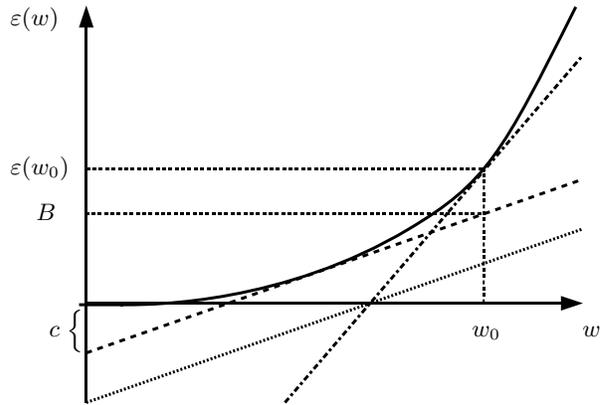}}
\put(0.,6){\mbox{$\varepsilon(w)$}}
\put(0.,3.7){\mbox{$\varepsilon(w_0)$}}
\put(0.4,3){\mbox{$B$}}
\put(0.6,1.2){\mbox{$c \; \Big\{$}}
\put(8.8,1.2){\mbox{$w$}}
\put(7.1,1.2){\mbox{$w_0$}}
\end{picture}
\caption{A schematic view of the estimation method. In order 
to estimate the convex function $\varepsilon(w)$ we consider 
linear affine functions below it. The dotted line 
corresponds to a general estimate as in Eq.~(\ref{affbound}), 
the dashed line to an estimate with the same slope $r$, but 
the smallest possible $c.$  This gives already the bound
$\varepsilon(w_0) \geq B = r \cdot w_0 -c.$ 
By varying the slope $r$ one arrives at the dashed-dotted 
line, which gives the best possible bound $\varepsilon(w_0).$
}
\label{fig:Legendre}
\end{figure}

The idea of our estimate is to characterize a convex function such as 
$\Emin:\Rl^n\to\Rl$ or the
entanglement measure $E$ itself as the supremum of all affine (i.e.,
linear+constant) functions below it. So let $r=(r_1,\ldots,r_n)$ and 
$w=(w_1,\ldots,w_n)$ be vectors, which we use to define the linear 
function $w\mapsto r\cdot w=\sum_kr_kw_k$, and consider bounds of 
the type
\begin{equation}
\label{affbound}
    \Emin(w)\geq r\cdot w -c
\end{equation}
for arbitrary $r$ and $c.$ Note that by definition of $\Emin$ 
this is the same as saying that $E(\rho)\geq r\cdot w-c$ for 
every $\rho$ giving the expectation values $w_k$ as in 
(\ref{Emin}). The constant $c$, which we try to choose as small 
as possible, hence needs to satisfy, for any $\rho$, the inequality
\begin{equation}
\label{cbelow}
    c\geq \sum_kr_k\tr(\rho \WW_k)\ - E(\rho),
\end{equation}
where we already inserted the condition $w_k=\tr(\rho \WW_k)$.
Obviously, the best choice of $c$ is the supremum of the right hand
side, which only depends on the operator $\WW=\sum_k r_k \WW_k$. Hence 
we can write
\begin{eqnarray}
\label{legendreE}
    c&=& \Eh\Bigl(\sum_k r_k \WW_k \Bigr) \quad\mbox{with}\\
       \label{Ehat}
    \Eh(\WW)&=&\sup_\rho\{\tr(\rho \WW)-E(\rho)\}.
\end{eqnarray}
Here (\ref{Ehat}) is just the definition of $\Eh$ as the Legendre
transform of the 
entanglement measure $E$. We now use the
formula (\ref{legendreE}) of the optimal constant $c$ in
(\ref{affbound}) to compute $\Emin$. As a convex function it is the
supremum over all affine functions below it, which are now
parameterized by the ``slopes'' $r$ (see also Fig.~\ref{fig:Legendre}). 
Hence we arrive at the main formula of this paper, characterizing the 
lower bound on $E$, obtainable from the measured expectations $w_k$:
\begin{equation}\label{EminLeg}
    \Emin(w)=\sup_r\textstyle\Bigl\{
           r\cdot w-\Eh\bigl(\sum_k r_k \WW_k\bigr)\Bigr\}.
\end{equation}
Once again this is a Legendre transform formula, saying that $\Emin$
is the Legendre transform of
$
\widehat\Emin(r)=\textstyle\Eh\bigl(\sum_kr_k \WW_k\bigr).
$

Of course, we want to apply formula (\ref{EminLeg}) mainly when
$n=1$, or at least, when $n$ is very small compared to the dimension
of the full space of hermitian operators. It does involve the
computation of two Legendre transforms: on the one hand, we have to
compute $\Eh$ from (\ref{Ehat}). For any choice of coefficients
$(r_1,\ldots,r_n)$ the computation of
$c=\Eh(\sum_kr_k \WW_k)$ already gives a partial solution to
our problem of giving a lower bound on $E(\rho)$ in terms of the
measured expectations, namely a best linear lower bound of the form
(\ref{affbound}). Optimizing over $r$ then gives the best overall
lower bound (\ref{EminLeg}) for which the Legendre transform has to
be taken over a low (i.e., $n$-) dimensional space only (see
Fig~\ref{fig:Legendre}). In any case the success of the method
depends on the possibility of efficiently computing $\Eh$. Clearly
this will depend on the entanglement measure $E$ and the witness $\WW$ 
chosen.
We will demonstrate now for important examples how the
computation can be done.

{\it Convex roof constructions ---} 
Many entanglement measures are defined 
by a standard extension process, which extends a function 
$\ket{\psi}\mapsto E(\ket\psi)$ defined only on pure entangled states to 
all mixed states, namely as
\begin{equation}\label{roof}
    E(\rho) = \inf_{p_i, \ket{\psi_i}} \sum_i p_i E(\ket{\psi_i}),
\end{equation}
where the $p_i$ are convex weights, and $\sum_i p_i \ketbra{\psi_i}
= \rho$. The convex roof (or ``convex hull'') is just the largest
convex function smaller than $E$ on pure states, and can therefore
be computed as the supremum of its affine lower bounds, i.e., once again as
a Legendre transform. $\Eh$ can then be simplified to a variational
problem over pure states only:
\begin{eqnarray}
\label{roofhat}
\Eh(\WW) 
&=& \sup_\rho
\Bigl\{ \tr(\rho \WW) - \inf_{p_i, \ket{\psi_i}}  \sum_i p_i E(\ket{\psi_i})\Bigr\}
\nonumber \\
&=& \sup_{p_i}
\sup_{\ket{\psi_i}}
\Bigl\{\sum_ip_i
            \bigl\{\bra{\psi_i}\WW\ket{\psi_i}-
             E(\ket{\psi_i}) \bigr\}
\Bigr\}
\nonumber\\
   &=&\sup_{\ket{\psi}}
            \bigl\{\bra{\psi}\WW\ket{\psi}-
             E(\ket{\psi}) \bigr\}.
\end{eqnarray}
Here, at the second equality, we converted the ``$-\inf$'' into a
``$\sup$'' and  substituted $\rho$ from the constraint
$\rho=\sum_ip_i\ketbra{\psi_i}$. The constraint then becomes
redundant, because the $\sup$ is taken over all values $\rho$, too.
The sup over the $p_i$ can furthermore be dropped, because convex
combinations of expressions of the form (\ref{roofhat}) cannot be
larger than the largest of these values. So in the end we can use
the Legendre formula (\ref{Ehat}) for $\Eh$, with the simplification
that we need only vary over pure states.

In many cases the variation can be simplified by varying first over
orbits of the local unitary group, i.e., to consider vectors
$\ket{\psi}=(U_1\otimes U_2)\ket{\phi}$ with $U_1,U_2$ unitary 
matrices and $\ket{\phi}$ fixed. Since, by definition, entanglement 
measures are invariant under such transformations, the second term in
(\ref{roofhat}) is independent of the $U_i$, so we can maximize the
first term separately. 

Consider, for example witness operators of the form 
$\WW=\alpha \eins - \ketbra{\chi},$ which is a typical 
form of witnesses. Then we have to maximize 
$|\braket{\chi}{(U_1\otimes U_2)\phi}|$. It
is easy to see that this maximum is attained, when
$\ket{\chi}$ and $(U_1\otimes U_2)\ket{\phi}$ have 
the same Schmidt basis, and the Schmidt coefficients 
are ordered in the same way (for a detailed proof see the 
appendix of Ref.~\cite{nlew}). Hence for a system
composed of two $d$-dimensional ones, we only need to vary 
over $d$ positive numbers with one normalization constraint
(rather than $d^2$ complex amplitudes in $\ket{\psi}$). 
In the examples shown below this reduces the computation
to a simple one parameter optimization. 

%
%

{\it Entanglement of formation ---}
The entanglement of formation $E_F$ is defined as the convex roof of the
function $E_F(\ket\psi)=S\bigl(\tr_2(\ketbra\psi)\bigr)$, the von
Neumann entropy of the reduced state \cite{eof}. It is one of the natural
figures of merit for experimental achievements in state preparation,
because it quantifies the entanglement (measured in singlet pairs)
which must be invested per realization of the state. In contrast,
measures like the distillable entanglement tell us about the
potential further uses of the state, which may be quite low, even if
the state is entanglement-expensive to make.

For small dimensions the direct computation of $\Eh_F$ along the
lines described above is feasible. For higher dimensions it is
convenient to solve (\ref{roofhat}) by an uphill iteration, which
will find a maximum efficiently.

To this end we rewrite the entropy term by the Gibbs variational
principle, i.e., as the Legendre transform of the free energy $F$
from statistical mechanics:
\begin{eqnarray}
\label{entGibbs}
S(\rho) &=& \inf_H\bigl\{\tr\rho H-F(H) \bigr\} =-\tr\rho\ln\rho , 
\\
\label{}
F(H)  &=&\inf_\rho\bigl\{\tr\rho H-S(\rho)\bigr\}=-\ln\tr(e^{-H}).
\end{eqnarray}
Here the first infimum is over all hermitian operators $H$, and the
second is over all density operators $\rho$. The first infimum is
attained for $H=-\ln\rho$, and the second one for $\rho=\exp(-H)/\tr(\exp(-H))$. 
We followed the conventions from statistical mechanics by using 
natural logarithms, but have set the inverse temperature $\beta=1$
\cite{remark2}.
Inserting (\ref{entGibbs})
into the entanglement term in (\ref{roofhat}) we get
\begin{equation}
\label{EhFsupsup}
\Eh_F(\WW)=\sup_{\ket{\psi}}\sup_H
\bigl\{ \bra\psi (\WW-H \otimes\idty) \ket\psi +F(H)   \bigr\},
\end{equation}
where the first supremum is over all unit vectors of the bipartite
system, and the second over all hermitian $H$ of the first system.
The point of this way of writing $\Eh_F$ is that the suprema over
these two variables obviously commute, and that when one of them is
fixed, the supremum (in fact, the absolute maximum) over the other
variable can be computed directly (without a search algorithm).
Indeed, for fixed $H$ (\ref{EhFsupsup}) requires $\ket{\psi}$ to be 
an eigenvector for the largest eigenvalue of
$(\WW-H\otimes\idty)$. On the other hand, when $\ket{\psi}$ is fixed, the
variation is exactly (\ref{entGibbs}) for the reduced density
operator $\rho_1$ of $\ket{\psi}$, which we know to be attained at
$H=-\ln\rho_1.$ Hence by alternating these steps, we
gain in every step, and get convergence to a  local maximum.
In the cases we have tried, the local maximum was always independent
of the starting point, giving strong support to the claim of having
found the global maximum. Therefore the algorithm is a useful tool
for finding the maximum. However, a guarantee cannot be given in
this algorithm, so in principle the resulting entanglement lower
bound (\ref{EminLeg}) could be too optimistic.

{\it Geometric measure of entanglement ---}
This measure is an entanglement monotone for 
multipartite systems \cite{wei}, defined via 
the convex roof construction and 
\be
E_G(\ket{\psi}) = 1- \sup_{\ket{\phi}=\ket{a}\ket{b}\ket{c}...} 
|\braket{\phi}{\psi}|^2
\ee
as one minus the maximal squared overlap with the fully separable 
states. For pure states, the geometric measure is a lower bound 
on the relative entropy and one can derive from it an upper bound 
on the number of states which can be discriminated perfectly by local 
operations and classical communication \cite{geoapp}.
We have then
\be
\Eh_G(\WW) = 
\sup_{\ket{\psi}}
\sup_{\ket{\phi}=\ket{a}\ket{b}\ket{c}...} 
\bigl\{
\bra{\psi}(\WW + \ketbra{\phi})\ket{\psi}-1
\bigr\}.
\label{geoleg}
\ee
To show how this optimization can be performed, let us assume for 
simplicity that we have three parties, i.\ e., $\ket{\phi} = \ket{abc}$.
If $\ket{a}$, $\ket{b}$ and $\ket{c}$ are fixed, we can perform the optimization
by taking $\ket{\psi}$ as an eigenvector corresponding to the maximal 
eigenvalue. If we fix $\ket{\psi}$ and two of the other 
vectors, e.\,g., $\ket{b}$ 
and $\ket{c}$, we have to find a vector $\ket{\tilde{a}}$ such that
$
\sup_{\ket{a}} |\braket{\psi}{abc}|^2 =: |\braket{\psi}{\tilde{a}bc}|^2.
$
If the Schmidt decomposition of $\ket{\psi}$ with respect to the 
$A|BC$ partition is given by 
$\ket{\psi}=\sum_i s_i \ket{\eta^{A}_i} \ket{\eta^{BC}_i},$ 
we have $|\braket{\psi}{abc}| = 
|\sum_i s_i \braket{\eta^{A}_i}{a} \braket{\eta^{BC}_i}{bc}|.$ 
This scalar product is maximal if the vectors are parallel.
So we set 
\be
\ket{\tilde a} = \NN \sum_j s_j\braket{\eta^{BC}_j}{bc}\ket{\eta^{A}_j},
\ee
where $\NN$ denotes a normalization. So this optimization can be iterated, 
as in the case of the entanglement of formation. Note that a similar 
iteration also delivers a method to calculate the geometric measure 
$E_G(\ket{\psi})$ for arbitrary pure states $\ket{\psi}.$

For special cases of witnesses, the Legendre transform can even be 
calculated analytically. Let us assume that the witness is of the 
form
$
r \WW = r (\alpha \eins - \ketbra{\chi}).
$
Here, we 
have already inserted the $r$ as it is used in Eq.~(\ref{EminLeg}).
If $r > 0,$ we choose in Eq.~(\ref{geoleg}) $\ket{\phi}$ orthogonal
to $\ket{\chi},$ resulting in $\Eh(r\WW)=r\alpha.$ If $r < 0,$ one 
can directly verify that we have to choose $\ket{\phi}$ as the 
state with the largest overlap with $\ket{\chi},$ which results 
in 
\be
\Eh_G (r\WW) =  
\frac{1-r}{2}+ 
\frac{1}{2} \sqrt{(1-r)^2+4rE_G(\ket{\chi})} + r \alpha -1.
\label{leggeoeinfach}
\ee
Hence $\Eh_G$ can be computed, provided $E_G(\ket{\chi})$ is 
known.

{\it Application to the experiment ---}
The experiment  in Ref.~\cite{mohamed} aimed at the production of the 
W-state
\be
\ket{W}=\frac{1}{\sqrt{3}}(\ket{001}+\ket{010}+\ket{100}).
\ee
For the entanglement verification, two witnesses have been used.
The witnesses and their mean values were given by \cite{mohamed}
\bea
&&\WW_1 = \frac{2}{3} \eins - \ketbra{W} , 
\;\;\;\;\;\;\;\;\;\;\;\;  
\mean{\WW_1}=-0.197 \pm 0.018,
\nonumber
\\
&&\WW_2 = \frac{1}{2} \eins - \ketbra{\psi^{\rm GHZ}} , 
\;\; 
\mean{\WW_2}=-0.139 \pm 0.030,
\nonumber
\eea
where $\ket{\psi^{\rm GHZ}}= (\ket{y^+ y^+ y^+}-\ket{y^- y^- y^-})/\sqrt{2} 
= {i}(\sqrt{3}\ket{W}-\ket{111})/2$ is a GHZ type state.

For the entanglement of formation, we consider 
the $A|BC$-bipartition, because 
of the symmetry the other bipartitions are equivalent. 
If we apply our theory on witnesses $\WW_1$ and $\WW_2$ 
separately, we get the bounds
$E_F^{(1)}(\rho)\geq 0.308 \pm 0.051$ from $\WW_1$  and 
$E_F^{(2)}(\rho)\geq 0.140 \pm 0.051$ from $\WW_2.$ If we 
use both witnesses at the same time, we get the  bound
\be
E_F^{(1,2)}(\rho)\geq 0.309 \pm 0.050.
\ee
For the geometric measure, using  
Eq.~(\ref{leggeoeinfach}) and the 
fact that $E_G(\ket{W})=5/9$ 
and  $E_G(\ket{\psi^{\rm GHZ}})=1/2$ \cite{wei}, 
we get 
the bounds
$E_G^{(1)}(\rho)\geq 0.199 \pm 0.022$ from $\WW_1$  and 
$E_G^{(2)}(\rho)\geq 0.019 \pm 0.010$ from $\WW_2.$ 
Using both witnesses simultaneously,  we obtain the bound
\be
E_G^{(1,2)}(\rho)\geq 0.209 \pm 0.023.
\ee
The fact that the bounds from $\WW_1$ are better than the ones 
obtained from $\WW_2$ stems from the fact that $\WW_1$ is by 
construction sensitive for detecting the W-state. If the W-state 
were produced perfectly, then the bound from $\WW_1$ would give
the exact value, since only the W state is compatible with 
$\mean{\WW_1}=-1/3.$ Naturally, the bounds using both witnesses 
are always better than the bound of the single witnesses alone, 
since more information on the state is available. In principle, 
one may still improve the bound by including all the measured 
coincidence probabilities from Ref.~\cite{mohamed}.

Along the same lines one can also investigate other experiments, 
where witnesses have been used \cite{witexp, mohamed}. In the 
exceptional cases where complete state tomography has been done 
\cite{hartmut}, one may, of course, also try other estimation methods.
Then it would be of great interest to compare these methods with 
our proposed one. 

{\it Conclusion ---}
We proposed a method to estimate entanglement measures
in experiments. To do so, we showed how entanglement 
witnesses can be used to obtain lower bounds on generic 
entanglement measures. We have explicitly demonstrated 
the calculations for the 
entanglement of formation and the geometric measure of entanglement.
Finally, we applied our results to experimental data, gaining new insights
into already performed experiments. Identifying witnesses, which are not only
capable to detect entanglement in noisy situations but deliver at the same 
time good estimates of entanglement measures is an interesting task for 
further study.

We thank H. J. Briegel, J. Eisert, A. Miyake, and K. Osterloh 
for valuable discussions. This work has been supported 
by the FWF, the DFG and the EU (OLAQI, PROSECCO, QUPRODIS, QICS, SCALA).


\begin{thebibliography}{12}
\bibitem{remark1} Taking the $\WW_i$ as witnesses
with $Tr(\rho\WW_i)<0$ for at least one $i$ guarantees 
that the bound on the entanglement measure will not 
be the trivial bound $E(\rho)\geq 0.$
\bibitem{remark2} Since $E_F(\ket{\psi})$ is defined 
with a $\log_2$ in the von Neumann entropy, one has 
to rescale the obtained value at the end.
\end{thebibliography}
\end{document}